\documentclass[12pt]{article}
\usepackage{mitpress}
\usepackage{amsmath}
\usepackage{graphicx}
\usepackage{amsfonts}
\usepackage{amssymb}
\newdimen\dummy
\dummy=\oddsidemargin
\begin{document}

\title{Classical, quantum and total correlations}
\author{L. Henderson$^{\ast}$ and V. Vedral$^{\ast\ast}$\\$^{\ast}$Department of Mathematics, University of Bristol, University Walk,
Bristol BS8 1TW\\$^{\ast\ast}$Optics Section, Blackett Laboratory, Imperial College, Prince
Consort Road, London SW7 2BZ}
\maketitle
\begin{abstract}
We discuss the problem of separating consistently the total correlations in a
bipartite quantum state into a quantum and a purely classical part. A measure
of classical correlations is proposed and its properties are explored.
\end{abstract}

In quantum information theory it is common to distinguish between purely
classical information, measured in bits, and quantum information, which is
measured in qubits. These differ in the channel resources required to
communicate them. Qubits may not be sent by a classical channel alone, but
must be sent either via a quantum channel which preserves coherence or by
teleportation through an entangled channel with two classical bits of
communication \cite{Bennett93}. In this context, one qubit is equivalent to
one unit of shared entanglement, or `e-bit', together with two classical bits.
Any bipartite quantum state may be used as a communication channel with some
degree of success, and so it is of interest to determine how to separate the
correlations it contains into a classical and an entangled part. A number of
measures of entanglement and of total correlations have been proposed in
recent years \cite{Smolin96, Bennett96, Vedral97PRL, Vedral98, Plenio98}.
However, it is still not clear how to quantify the purely classical part of
the total bipartite correlations. In this paper we propose a possible measure
of classical correlations and investigate its properties.

We first review the existing measures of entangled and total correlations. In
classical information theory, the Shannon entropy, $H(X)\equiv H(p)=-\sum
_{i}p_{i}\log p_{i}$, is used to quantify the information in a source, $X$,
that produces messages $x_{i}$ with probabilities $p_{i}$ \cite{Shannon49,
Cover91}. The relative entropy is a useful measure of the closeness of two
probability distributions $\{p_{i}\}$ and $\{q_{i}\}$ from the same source
$X$. The relative entropy of $\{p_{i}\}$ to $\{q_{i}\}$ is defined as
$H(p||q)=\sum_{i}p_{i}\log\frac{p_{i}}{q_{i}}$. Correlations between two
different random variables $X$ and $Y$ are measured by the mutual information,
$H(X:Y)=H(X)+H(Y)-H(X,Y)$, where $H(X,Y)=-\sum_{i,j}p_{ij}\log p_{ij}$ is the
joint entropy and $p_{ij}$ is the probability of outcomes $x_{i}$ and $y_{j}$
both occurring. The mutual information measures how much information $X$ and
$Y$ have in common. It may also be defined as a special case of the relative
entropy, since it is a measure of how distinguishable a joint probability
distribution $p_{ij}$ is from the completely uncorrelated pair of
distributions $p_{i}p_{j}$, $H(p_{ij}||p_{i}p_{j})=H(p_{i})+H(p_{j}%
)-H(p_{ij})$.

In a quantum context, the results of a measurement $\{E_{y}\}$ on a state
represented by a density matrix, $\rho$, comprise a probability distribution
$p_{y}=Tr(E_{y}\rho)$. The Von Neumann entropy is a way of measuring the
information in a quantum state by taking the entropy of the probability
distribution generated from the state $\rho$ by a projective measurement onto
the state's eigenvectors \cite{Neumann55}. It is defined as $S(\rho
)=-Tr(\rho\log\rho)=H(\lambda)$, where $\lambda=\{\lambda_{i}\}$ are the
eigenvalues of the state. The classical relative entropy and classical mutual
information also have analogues in the quantum domain. The quantum relative
entropy of a state $\rho$ with respect to another state $\sigma$ is defined as
$S(\rho||\sigma)=-S(\rho)-Tr(\rho\log\sigma)$. The joint entropy $S(\rho
_{AB})$ for a composite system $\rho_{AB}$ with two subsystems $A$ and $B$ is
given by $S(\rho_{AB})=-Tr(\rho_{AB}\log\rho_{AB})$ and the Von Neumann mutual
information between the two subsystems is defined as
\[
I(\rho_{A:B})=S(\rho_{A})+S(\rho_{B})-S(\rho_{AB})
\]
As in the classical case, the mutual information is the relative entropy
between $\rho_{AB}$ and $\rho_{A}\otimes\rho_{B}$. The mutual information is
usually used to measure the total correlations between the two subsystems of a
bipartite quantum system.

The entanglement of a bipartite quantum state $\rho_{AB}$ may be quantified by
how distinguishable it is from the `nearest' separable state, as measured by
the relative entropy. Relative entropy of entanglement, defined as
\[
E_{RE}(\rho_{AB})=\min_{\sigma_{AB}\in D}S(\rho_{AB}||\sigma_{AB})
\]
has been shown to be a useful measure of entanglement ($D$ is the set of all
separable or disentangled states) \cite{Vedral97PRL, Vedral98}. Note that
$E_{RE}(\rho_{AB})\leq I(\rho_{A:B})$, by definition of $E_{RE}(\rho_{AB})$,
since the mutual information is also the relative entropy between $\rho_{AB}$
and a completely disentangled state, $I=S(\rho_{AB}||\rho_{A}\otimes\rho_{B})$
and so must be higher than the minimum over all disentangled states.

Another way to measure the entanglement of a bipartite quantum state is to
consider the process of formation of an ensemble of entangled states
\cite{Smolin96}. The ensemble is first prepared locally by Alice, then one
subsystem is compressed using Schumacher compression \cite{Schumacher95} and
sent to Bob by teleportation. The entanglement of formation is then the amount
of entanglement required for the teleportation. For pure states this is given
by the compression efficiency, $E_{F}(\rho_{AB})=S(\rho_{B})$. For mixed
states, the entanglement of formation is $E_{F}(\rho_{AB})=\min\sum_{i}%
p_{i}S(\rho_{B}^{i})\leq S(\rho_{B})$, where the minimum is taken over all
decompositions of the mixed state. However, teleportation which requires only
this much entanglement must be accompanied by classical communication of
information about the decomposition of the mixed state \cite{Henderson00}. The
information in the subsystem $S(\rho_{B})$ is thus split into a classical part
and a quantum part. The classical part may be transmitted by a classical
channel, but the quantum part requires entanglement and is sent by teleportation.

There has been some work on the general problem of splitting information in a
particular quantum state into a classical and a quantum part \cite{Bennett94}.
Consider performing a general measurement on the state,  $A_{i}^{\dagger}%
A_{i}$, such that $\rho_{B}^{i}=\frac{A_{i}\rho_{B}A_{i}^{\dagger}}%
{tr(A_{i}\rho_{B}A_{i}^{\dagger})}$. The final state of subsystem $B$ is then
$\sum_{i}A_{i}\rho_{B}A_{i}^{\dagger}=\sum_{i}p_{i}\rho_{B}^{i}$. The entropy
of the residual states is $\sum_{i}p_{i}S(\rho_{B}^{i})$. The classical
information obtained by measuring outcomes $i$ with probabilities $p_{i}$ is
$H(p)$. If the states $\rho_{B}^{i}$ have support on orthogonal subspaces,
then the entropy of the final state is the sum of the residual entropy and the
classical information $S(\sum_{i}p_{i}\rho_{B}^{i})=H(p)+\sum_{i}p_{i}%
S(\rho_{B}^{i})$. It has been shown\textit{ }that the state $\rho_{B}=\sum
_{i}p_{i}\rho_{B}^{i}$ can be reconstructed with arbitrarily high fidelity
from the classical measurement outcomes and the residual states if and only if
the residual states $\rho_{B}^{i}$ are on orthogonal subspaces
\cite{Bennett94}. We see then that the information in a quantum state may be
split into a quantum and a classical part.

We now ask how this can be done for correlations between two subsystems. We
would like a way to measure the classical correlations between two subsystems.
We first suggest four reasonable properties we should expect a measure of
classical correlations, $C$, to satisfy.

\begin{enumerate}
\item $C=0$ for $\rho=\rho_{A}\otimes\rho_{B}$. This requires that product
states are not correlated.

\item $C$ is invariant under local unitary transformations. This is because
any change of basis should not affect the correlation between two subsystems.

\item $C$ is non-increasing under local operations. If the two subsystems
evolve \emph{independently} then the correlation between them cannot increase.

\item $C=S(\rho_{A})=S(\rho_{B})$ for pure states. This is a natural
requirement, as we will see below.
\end{enumerate}

Note that $(2)$ and $(4)$ are also required of a measure of entanglement. If
classical communication were added to $(3)$, it would be identical to the
corresponding condition for entanglement measures. However, if classical
communication is allowed, then the classical correlations could increase as
well as decrease, which is not satisfactory. It is also natural that the
measure $C$ should be symmetric under interchange of the subsystems $A$ and
$B$. This is because it should quantify the correlation between subsystems
rather than a property of either subsystem. However, we do not include this as
a separate constraint as it is not clear that this condition is independent
from $(1)-(4)$.

We now suggest a measure which satisfies these properties. The proposed
measure is:%
\begin{equation}
C_{B}(\rho_{AB})=\max_{B_{i}^{\dagger}B_{i}}S(\rho_{A})-\sum_{i}p_{i}%
S(\rho_{A}^{i})\label{eq:cb}%
\end{equation}
where $B_{i}^{\dagger}B_{i}$ is a POVM performed on the subsystem B and
$\rho_{A}^{i}=tr_{B}(B_{i}\rho_{AB}B_{i}^{\dagger})/tr_{AB}(B_{i}\rho
_{AB}B_{i}^{\dagger})$ is the remaining state of $A$ after obtaining the
outcome $i$ on $B$. Alternatively,%
\begin{equation}
C_{A}(\rho_{AB})=\max_{A_{i}^{\dagger}A_{i}}S(\rho_{B})-\sum_{i}p_{i}%
S(\rho_{B}^{i})\label{eq:ca}%
\end{equation}
if the measurement is performed on subsystem $A$ instead of on $B$. Clearly
$C_{A}(\rho_{AB})=C_{B}(\rho_{AB})$ for all states $\rho_{AB}$ such that
$\rho_{A}=\rho_{B}$. It remains an open question whether this is true in
general. The measure is a natural generalisation of the classical mutual
information, which is the difference in uncertainty about the subsystem $B$
($A$) before and after a measurement on the correlated subsystem $A$ ($B$),
$H(A:B)=H(B)-H(B|A)$. Similarly, Eq.s (\ref{eq:cb}) and (\ref{eq:ca})
represent the difference in Von Neumann entropy before and after the
measurement. Note the similarity of the definition to the Holevo bound which
measures the capacity of quantum states for classical communication
\cite{Holevo73}.

The following example provides an illustration. Consider a bipartite separable
state of the form
\[
\rho_{AB}=\sum_{i}p_{i}\left|  i\right\rangle \langle i|_{A}\otimes\rho
_{B}^{i}
\]
where $\{\left|  i\right\rangle \}$ are orthonormal states of subsystem $A$.
Clearly the entanglement of this state is zero. The best measurement that
Alice can make to gain information about Bob's subsystem is a projective
measurement onto the states $\{\left|  i\right\rangle \}$ of subsystem $A$.
Therefore the classical correlations are given by
\[
C_{A}(\rho_{AB})=S(\rho_{B})-\sum_{i}p_{i}S(\rho_{B}^{i})
\]
For this state, the mutual information is also given by
\[
I(\rho_{A:B})=S(\rho_{B})-\sum_{i}p_{i}S(\rho_{B}^{i})
\]
This is to be expected since there are no entangled correlations and so the
total correlations between $A$ and $B$ should be equal to the classical correlations.

We now investigate the properties of the quantities in Eq.s (\ref{eq:cb}) and
(\ref{eq:ca}). Property (1) above is clearly satisfied, since the state of
subsystem $B$ corresponding to any measurement result $i$ on subsystem $A$ is
still $\rho_{B}$ for a product state. In fact, $C(\rho_{AB})=0$ if and only if
$\rho_{AB}=\rho_{A}\otimes\rho_{B}$. Property (2) is satisfied since the Von
Neumann entropy is invariant under local unitary transformations. Property (4)
is also satisfied, since for pure states $C_{A}(\rho_{AB})=S(\rho_{A})$
($C_{B}(\rho_{AB})=S(\rho_{B})=S(\rho_{A})=C_{A}(\rho_{AB})$) can always be
achieved by local projection onto the Schmidt basis. Therefore for pure states
$E(\rho_{AB})=C(\rho_{AB})$ and $I(\rho_{A:B})=2E(\rho_{AB})=2C(\rho_{AB})$
(here $E(\rho_{AB})$ may be either $E_{RE}(\rho_{AB})$ or $E_{F}(\rho_{AB})$
since these measures coincide for pure states). The most important property
required of a measure of classical correlations is that it is non-increasing
under local operations (property (3)). We now show that this property is
satisfied by the proposed measure.

\textbf{Property (3):} The measure $C_{A}$ ($C_{B}$) is non-increasing under
local operations.

\textbf{Proof:} Let $\{A_{i}^{\dagger}A_{i}:\sum_{i}A_{i}^{\dagger}A_{i}=I\}$
be the POVM which maximises $C_{A}=\max_{A_{i}^{\dagger}A_{i}}S(\rho_{B}%
)-\sum_{i}p_{i}S(\rho_{B}^{i})=\max_{A_{i}^{\dagger}A_{i}}\sum_{i}p_{i}%
S(\rho_{B}^{i}||\rho_{B})$.

a) Consider a local operation $\phi_{A}$ on subsystem $A$. This may be
regarded as part of the POVM on $A$ so $C_{A}$, being a maximum is not affected.

b) Now take a local operation $\phi_{B}$ on subsystem $B$. Then by the
property that the relative entropy does not increase under local operations,
$\sum_{i}p_{i}S(\rho_{B}^{i}||\rho_{B})\geq\sum_{i}p_{i}S(\phi_{B}(\rho
_{B}^{i})||\phi_{B}(\rho_{B}))$ \cite{Lindblad75}. Therefore $C_{A}$ does not
increase under local operations.

We now consider the relations between the classical, total and entangled
correlations in some simple cases. These raise some interesting general questions.

\begin{figure}
[ptb]
\begin{center}
\includegraphics[
height=3.1618in,
width=3.8683in
]%
{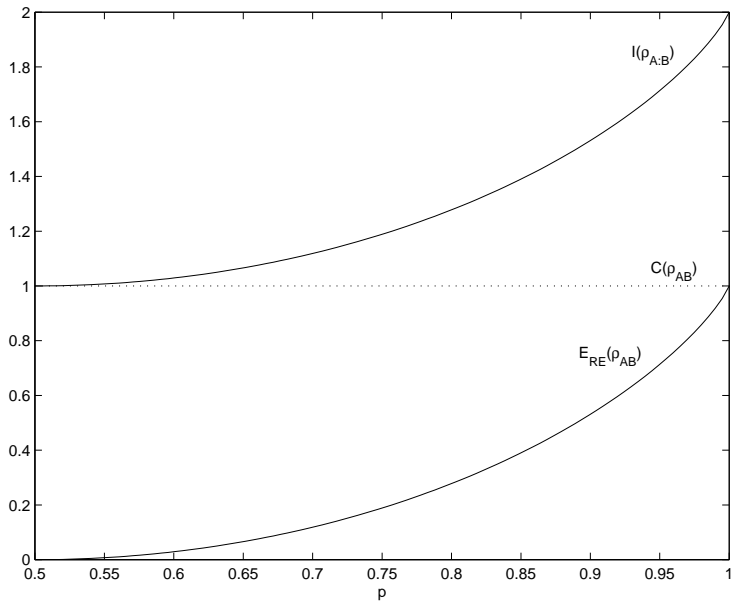}%
\caption{Correlations for a mixture of two Bell states, $\rho_{AB}=p\left|
\phi^{+}\right\rangle \langle\phi^{+}|+(1-p)\left|  \phi^{-}\right\rangle
\langle\phi^{-}|$, as a function of $p$. }%
\label{fig:twobell}%
\end{center}
\end{figure}

First, consider a maximally entangled pure state, $|\phi^{+}\rangle\langle
\phi^{+}|$, and the family of states that interpolate between it and its
completely decohered state $\left|  00\right\rangle \langle00|+\left|
11\right\rangle \langle11|$. These are states of the form
\[
\rho_{AB}=p\left|  \phi^{+}\right\rangle \langle\phi^{+}|+(1-p)\left|
\phi^{-}\right\rangle \langle\phi^{-}|
\]
where $\frac{1}{2}\leq p\leq1$. The mutual information as a function of $p$
is\ $I(\rho_{A:B})=2+p\log p+(1-p)\log(1-p)$. The entanglement is $E_{RE}%
(\rho_{AB})=1+p\log p+(1-p)\log(1-p)$ \cite{Vedral98}. However the classical
correlations remain constant at $C_{A}(\rho_{AB})=C_{B}(\rho_{AB})=C(\rho
_{AB})=1$. This is achieved by a projective measurement onto $\{\left|
0\right\rangle \langle0|,\left|  1\right\rangle \langle1|\}$, and must be the
maximum because $C$ cannot exceed one. For this example, the total
correlations are just the sum of the entangled and the classical correlations,
$I(\rho_{A:B})=E_{RE}(\rho_{AB})+C(\rho_{AB})$, see Fig. (\ref{fig:twobell}).
%TCIMACRO{\FRAME{ftbpFU}{3.8683in}{3.1618in}{0pt}{\Qcb{Correlations for a
%mixture of two Bell states, $\rho_{AB}=p\left|  \phi^{+}\right\rangle
%\langle\phi^{+}|+(1-p)\left|  \phi^{-}\right\rangle \langle\phi^{-}|$, as a
%function of $p$. }}{\Qlb{fig:twobell}}{figtwobell.eps}%
%width 3.8683in;  height 3.1618in;  depth 0pt;  original-width 6.5994in;
%original-height 5.386in;  cropleft "0";  croptop "1";  cropright "1";
%cropbottom "0";  filename 'figtwobell.eps';file-properties "XNPEU";}}}%
%BeginExpansion
%EndExpansion

\begin{figure}
[ptb]
\begin{center}
\includegraphics[
height=3.1618in,
width=3.8683in
]%
{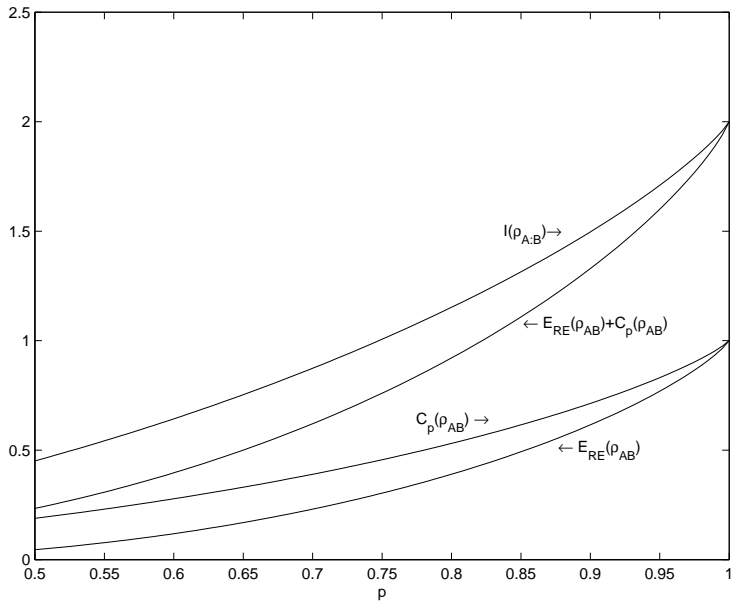}%
\caption{Correlations for a Werner state, $\rho_{AB}=p\left|  \phi
^{+}\right\rangle \langle\phi^{+}|+\frac{1-p}{4}I$, as a function of $p$.}%
\label{fig:Werner}%
\end{center}
\end{figure}

We now consider a Werner state of the form%
\[
\rho_{AB}=p\left|  \phi^{+}\right\rangle \langle\phi^{+}|+\frac{1-p}{4}I
\]
with $\frac{1}{2}\leq p\leq1$. The mutual information is given by
$I(\rho_{A:B})=2+f\log f+(1-f)\log(\frac{1-f}{3})$, where $f=\frac{3p+1}{4}$.
The relative entropy of entanglement is $E_{RE}(\rho_{AB})=1+f\log
f+(1-f)\log(1-f)$ \cite{Vedral98}. The state is symmetric under interchange of
subsystems $A$ and $B$, so $C_{A}(\rho_{AB})=C_{B}(\rho_{AB})\equiv
C(\rho_{AB})$. Any orthogonal projection produces the same value for the
classical correlations. We call this quantity $C_{p}(\rho_{AB})$. Clearly
$C_{p}(\rho_{AB})\leq C(\rho_{AB})$. These quantities are plotted in Fig.
(\ref{fig:Werner}).%
%TCIMACRO{\FRAME{ftbpFU}{3.8683in}{3.1618in}{0pt}{\Qcb{Correlations for a
%Werner state, $\rho_{AB}=p\left|  \phi^{+}\right\rangle \langle\phi^{+}%
%|+\frac{1-p}{4}I$, as a function of $p$.}}{\Qlb{fig:Werner}}{figwerner.eps}%
%width 3.8683in;  height 3.1618in;  depth 0pt;  original-width 6.5994in;
%original-height 5.386in;  cropleft "0";  croptop "1";  cropright "1";
%cropbottom "0";  filename 'figwerner.eps';file-properties "XNPEU";}}}%
%BeginExpansion
%EndExpansion

\begin{figure}
[ptb]
\begin{center}
\includegraphics[
height=3.1618in,
width=3.9159in
]%
{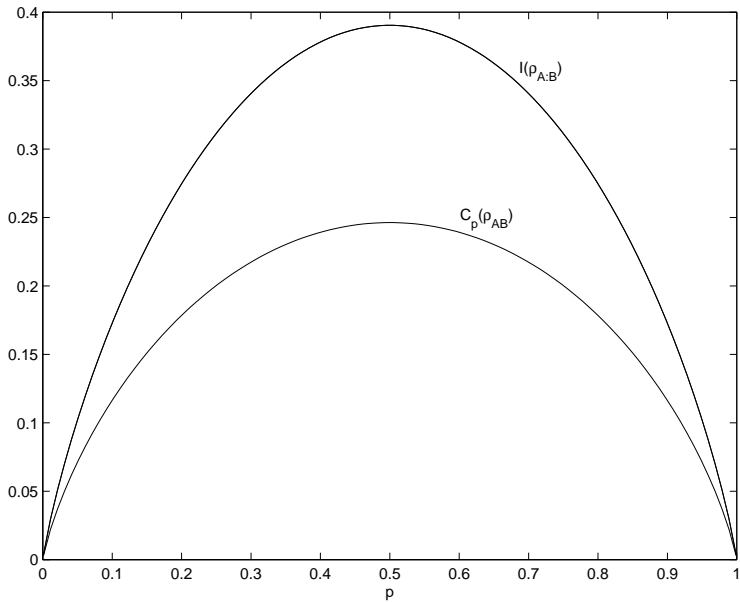}%
\caption{Correlations for the separable state, $\rho_{AB}=p|0\rangle
|0\rangle\langle0|\langle0|+(1-p)|+\rangle|+\rangle\langle+|\langle+|$, as a
function of $p$.}%
\label{fig:nonorth}%
\end{center}
\end{figure}

Consider now a state of the form
\[
\rho_{AB}=p|0\rangle|0\rangle\langle0|\langle0|+(1-p)|+\rangle|+\rangle
\langle+|\langle+|
\]
Again, the state is symmetrical with regard to $A$ and $B$, so $C_{A}%
(\rho_{AB})=C_{B}(\rho_{AB})\equiv C(\rho_{AB})$. This state provides a simple
example where the states on both sides are non-orthogonal. In this case, the
optimal single-shot measurement for distinguishing the two states $|0\rangle$
and $|+\rangle$ with respect to probability of error is known
\cite{Helstrom76}. However interestingly it is not the measurement which
optimises the classical correlations. We optimise over all orthogonal
measurements and call the resulting quantity $C_{p}(\rho_{AB})$. This is
plotted in Fig. (\ref{fig:nonorth}), together with the mutual information.%
%TCIMACRO{\FRAME{ftbpFU}{3.9159in}{3.1618in}{0pt}{\Qcb{Correlations for the
%separable state, $\rho_{AB}=p|0\rangle|0\rangle\langle0|\langle
%0|+(1-p)|+\rangle|+\rangle\langle+|\langle+|$, as a function of $p$.}%
%}{\Qlb{fig:nonorth}}{fignonorth.eps}{\special{ language "Scientific Word";
%type "GRAPHIC";  maintain-aspect-ratio TRUE;  display "USEDEF";
%valid_file "F";  width 3.9159in;  height 3.1618in;  depth 0pt;
%original-width 6.6815in;  original-height 5.386in;  cropleft "0";
%filename 'fignonorth.eps';file-properties "XNPEU";}}}%
%BeginExpansion
%EndExpansion

In these last two examples, we see that $C_{p}(\rho_{AB})+E_{RE}(\rho
_{AB})<I(\rho_{A:B})$. If the classical correlations are maximised by an
orthogonal measurement on one subsystem, ($C_{p}(\rho_{AB})=C(\rho_{AB})$),
the classical and entangled correlations do not account for all the total
correlations. This may indicate that the quantum mutual information is not the
best quantity for measuring total correlations, or that correlations are
simply not additive in this sense. However, $C_{p}(\rho_{AB})$ may not
coincide with $C(\rho_{AB})$. It is also possible that an asymptotic
measurement on many copies of the state would achieve a higher value for the
classical correlations than measurements on a single copy. This is because the
classical correlations are super-additive, $C(\rho\otimes\rho)\geq2C(\rho)$.
It is interesting to note that on the other hand, entangled correlations, as
measured by $E_{RE}$ or $E_{F}$, are subadditive, $E(\rho\otimes\rho
)\leq2E(\rho)$, and total correlations, measured by the mutual information,
are additive $I(\rho\otimes\rho)=2I(\rho)$.

A number of interesting questions are raised about the general relations
between $I$, $E$ and $C$. We do not know whether the sum of the two types of
correlations is generally greater than, less than or equal to the total
correlations, when asymptotic measurements are taken into account. For mixed
states, we saw that it need no longer be true that $E(\rho_{AB})=C(\rho_{AB}%
)$, as it is for pure states. This raises the question of whether $E(\rho
_{AB})=C(\rho_{AB})$ if and only if $\rho_{AB}$ is pure. In our examples we
found $E(\rho_{AB})\leq C(\rho_{AB})$, and we conjecture that this is
generally true. We know that $E_{RE}(\rho_{AB})\leq I(\rho_{A:B})$. Is it also
true that $C(\rho_{AB})\leq I(\rho_{A:B})$ in general?

Another possible measure of classical correlations could be based on the
relative entropy, just as measures of total and entangled correlations are
both relative entropies, $I(\rho_{A:B})=S(\rho_{AB}||\rho_{A}\otimes\rho_{B}%
)$, and $E(\rho_{A:B})=\min_{\sigma_{AB}\in D}S(\rho_{AB}||\sigma_{AB})$
\cite{Vedral97PRL, Vedral98}. Classical correlations could then be given by
the relative entropy between the closest separable state, $\sigma_{AB}^{\ast}%
$, and the product state $\rho_{A}\otimes\rho_{B}$, $C_{RE}=$ $S(\sigma
_{AB}^{\ast}||\rho_{A}\otimes\rho_{B})$. For the example of a mixture of two
Bell states, $C_{RE}(\rho_{AB})$ coincides with $C(\rho_{AB})=1$. For the
separable state $\rho_{AB}=p|0\rangle|0\rangle\langle0|\langle
0|+(1-p)|+\rangle|+\rangle\langle+|\langle+|$, $C_{RE}(\rho_{AB})=I(\rho
_{A:B})$, which makes sense intuitively since there is no entanglement.
However, for Werner states, the relative entropy of classical correlations
remains constant at $C_{RE}(\rho_{AB})=0.2075$. Therefore for low values of
$p$, $C_{RE}(\rho_{AB})>E_{RE}(\rho_{AB})$, whereas for high values,
$C_{RE}(\rho_{AB})<E_{RE}(\rho_{AB})$. In general $I(\rho_{A:B})>C_{RE}%
(\rho_{AB})+E_{RE}(\rho_{AB})$, so that the two types of correlations do not
sum to the total. It also remains to be proved whether $C_{RE}$ is
non-increasing under local operations.

In this paper we have raised the question of how to quantify the purely
classical part of a correlated quantum system, and we have suggested a
potential candidate for a measure which satisfies the most important property
of being non-increasing under local operations. A number of interesting open
questions about the relationship between measures of classical, entangled and
total correlations have been raised. It is hoped that a quantitative
understanding of the different types of correlations would aid our
understanding of protocols involving manipulation of entanglement and
classical information. In particular it should shed some light on the
conversion from entanglement to classical information which occurs in the
process of quantum measurement.

\section{Acknowledgments}

VV would like to thank G. Lindblad and N. Schou for useful discussions and the
EPSRC and European Commission (EQUIP grant) for financial support. Thank you
to E. Galvao for reading and commenting on the manuscript.

\bibliographystyle{prsty}
\bibliography{two}
\end{document}